\documentclass[aps,superscriptaddress,prd,
reprint,
floatfix, 
nofootinbib,
amsmath,amssymb,amsfonts,longbibliography]{revtex4-2}

\usepackage[pdftex]{graphicx}
\usepackage{hyperref}
\hypersetup{
setpagesize=false,
bookmarksnumbered=true,
bookmarksopen=true,%
colorlinks=true,%
linkcolor=red,
citecolor=blue,
}
\usepackage{float,color}
\usepackage{bm}
\usepackage{rsfso}
\usepackage{physics}
\usepackage{mathrsfs} 
\usepackage{comment}

\newcommand{\annob}[1]{\textcolor{blue}{#1}}

\newcommand{\two}{{I\hspace{-1.2pt}I}}
\newcommand{\three}{{I\hspace{-1.2pt}I\hspace{-1.2pt}I}}
\newcommand{\four}{{I\hspace{-1.2pt}V}}

\begin{document}
\title{Hawking radiation in quantum Hall system with an expanding edge:\\ application of anomaly method }
\author{Riku Yoshimoto }
\email{yoshimoto.riku.d1@s.mail.nagoya-u.ac.jp}
\author{Yasusada Nambu}
\email{nambu.yasusada.e5@f.mail.nagoya-u.ac.jp}
\affiliation{Department of Physics, Graduate School of Science, Nagoya University, Chikusa, Nagoya 464-8602, Japan}

\date{\today}

\begin{abstract}
The relationship between gravitational anomalies and Hawking radiation of black holes was revealed by Wilczek and Robinson. In this study, we apply their method to an analogue de Sitter spacetime in the quantum Hall system with an expanding edge. Because this system is chiral,  there is no need to impose the condition of ingoing modes near the horizon, which was assumed in the original method. Moreover, this system is structured so that the de Sitter space is sandwiched between two flat spaces, and although the effects of the anomaly would not appear in an ordinal de Sitter spacetime, they manifest themselves as boundary conditions between the de Sitter and the flat regions. By performing calculations under these boundary conditions, we obtain the flux of Hawking radiation in the outer flat region with the Gibbons-Hawking temperature of the de Sitter horizon.
\end{abstract}

\maketitle

\section{Introduction}
Hawking radiation has been studied extensively as a quantum property of black holes.  Because of its low temperature, observing it from actual black holes in the universe is extremely challenging. However, in condensed matter systems, it is possible to construct analog models with properties such as black hole horizons and cosmological horizons. Among them, the analog de Sitter spacetime with quantum Hall (QH) systems is discussed in \cite{Hotta2022b, Nambu2023a}. The usual experiments on QH systems are performed in a static situation where the edge is not time-varying, but by making it time-varying, the excitations moving along the expanding edge can be a simulator of the chiral scalar field in a (1+1)-dimensional curved spacetime. Also, in \cite{Hotta2022b, Nambu2023a}, the expanding edge can reproduce a two-dimensional de Sitter universe which has the spacetime structure similar to that of a black hole formation via gravitational collapse. It was shown that the entanglement between two spatial regions decreases due to Hawking radiation from the de Sitter region, and when the detection region is sufficiently large compared to the Hubble length of the de Sitter region, the two regions become separable, and only classical correlations remain as well as the inflationary universe.

In this paper, we focus on the chiral properties of edge excitations rather than on quantum correlations in the QH system. This implies the presence of chiral scalar fields in the analog de Sitter spacetime. In such a case, it is known that there is a gravitational anomaly~\cite{Alvarez-Gaume:1983ihn,Bertlmann:1996xk,Bertlmann:2000da}. Hawking radiation has been studied using various approaches beyond Hawking's original calculation~\cite{Hawking:1975vcx}. These include approaches using Euclidean quantum gravity by Hawking and Gibbons~\cite{Gibbons:1976ue}, calculations using the trace anomaly by Christensen and Fulling~\cite{Christensen:1977jc,Birrell1984}, and calculations based on quantum tunneling~\cite{Parikh:1999mf}. Robinson and Wilczek provided a new calculation of Hawking radiation using gravitational anomaly~\cite{Robinson:2005pd}.  Classical ingoing modes do not affect the dynamics of the external region near the black hole horizon. Thus, when they are ignored, the theory becomes chiral and a gravitational anomaly appears. Since the original theory is general covariant, this anomaly needs to be canceled by quantum effects, leading to the interpretation of obtaining Hawking radiation. This method has been applied not only to Schwarzschild black holes, but also to Kerr black holes, Reissner-Nordström black holes, and others~\cite{Iso:2006ut,Iso:2006wa,Murata:2006pt,Jiang:2007wj}. 

However, it is known that this method provides incorrect results for the Unruh effect and cosmological particle creation in the de Sitter spacetime~\cite{Akhmedova:2010zz,Zampeli:2013lka}. In this study, we evaluated the flux of Hawking radiation from the de Sitter horizon in the analog de Sitter spacetime  using the anomaly method. In the usual de Sitter spacetime, covariant gravitational anomaly method fails because the anomaly term becomes zero in de Sitter spacetime; because the spacetime curvature is constant and provides no terms to cancel out the anomaly. However, in the present QH system, the anomaly term does not become zero at the boundary. Therefore, the boundary conditions between the Minkowski and de Sitter regions are determined by the anomaly equation, and we can obtain the flux with Gibbons-Hawking temperature. 

The structure of the paper is as follows.
In Section \two, we review the coordinate transformations of each region in the expanding edge of QH systems introduced in \cite{Nambu2023a,Hotta2022b} and organize the coordinates to be used in later sections. In Section \three, we calculate the expectation value of the energy-momentum tensor (EMT) in the analog de Sitter spacetime introduced in Section {\two} based on the anomalous conservation law and the boundary conditions obtained from it. Section {\four} is devoted to summary and conclusions.

\newpage
\section{Quantum Hall system with expanding edge}
\begin{figure}[H]
\centering
\includegraphics[width=1\linewidth]{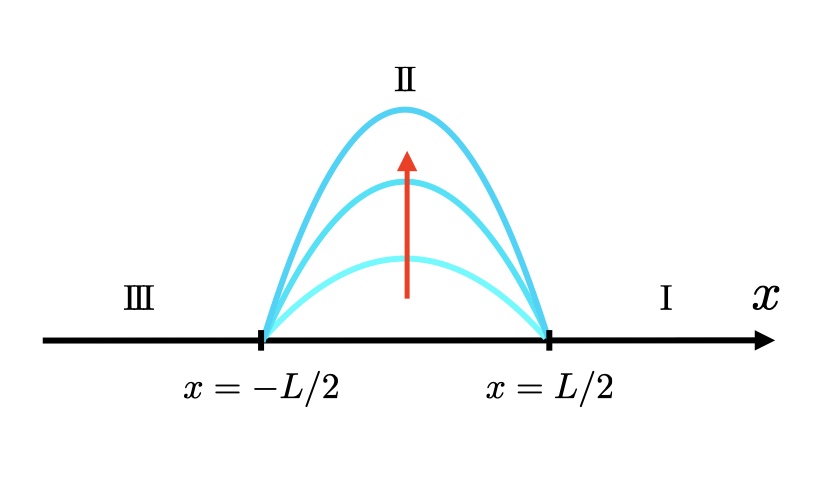}
\caption{Schematic picture of an expanding edge in QH systems. Regions I and {\three} are static and region {\two} is the expanding region.}
\label{fig:fig1}
\end{figure}

In this section, we review the coordinate setup for the expanding edge of the QH system~\cite{Nambu2023a,Hotta2022b}. We consider the edge of the QH system that has left-moving edge excitation. Then we divide the entire edge into three regions as shown in Fig.~\ref{fig:fig1}; Region I ($x>L/2$) and region {\three} ($x<-L/2$) are static, while region {\two} ($-L/2<x<L/2$) expands with time.
\noindent
The analog metric of this system is written as
\begin{align}
&\text{Region {\three}:}~x_\text{\three}\leq -L/2,\quad ds^2=-dx_\text{\three}^+\,dx_\text{\three}^-,\notag \\
&\text{Region {\two}:}~|x|\leq L/2,\quad ds^2=-e^{2\Theta(t)}dx^+dx^-, \notag \\
&\text{Region I:}~ L/2\leq x_\text{I},\quad ds^2=-dx_\text{I}^+dx_\text{I}^-,\notag 
\end{align}
where $e^{\Theta(t)}$ is the scale factor in region {\two} and $x^\pm_\text{I,\three}=t_\text{I,\three}\pm x_\text{I,\three},~x^{\pm}=t\pm x$ are  null coordinates in each region.

The conformal flat coordinates in region {\two} become discontinuous at the boundary $x=\pm L/2$, so it is not possible to smoothly extend them to regions I and {\three}. However, it is possible to introduce smooth coordinates $x_{\rm{I}}$ ($x_\text{\three}$) that cover both regions I and {\two} (or {\two} and {\three}), while keeping $x_{\rm{I}}=L/2$ ($x_\text{\three}=-L/2$).

\subsection{Region {\two}  and {\three} }
First, we perform the coordinate transformation between the regions {\two} and {\three} and express the metric in the region {\two} with $x_\text{\three}^{\pm}$. The matching condition at the boundary $x_\text{\three}=-L/2$ is
\begin{equation}
x_\text{\three}^{+}\left[t-L/2\right]-x_\text{\three}^{-}\left[t+L/2\right]=-L.
\end{equation}
By taking a derivative, we obtain
\begin{equation}
\frac{d x_\text{\three}^{+}}{dx^+}\left[t-L/2\right]=\frac{d x_\text{\three}^{-}}{dx^-}\left[t+L/2\right].
\end{equation}
Since the conformal factor is unity at $x=x_\text{\three}=-L/2$, we obtain
\begin{equation}
\frac{d x_\text{\three}^{+}}{dx^+}\left[t-L/2\right]=e^{\Theta (t)}.
\end{equation}
By shifting the argument of this function,
\begin{equation}
\frac{d x_\text{\three}^{+}}{d x^+}\left[x^+\right]=e^{\Theta (x^++L/2)},
\end{equation}
and one finds
\begin{align}
x_\text{\three}^+\left[x^+\right]&=\int^{x^+}_{0} dy\, e^{\Theta (y+L/2)}
=\int^{x^++L/2}_{L/2} dy\, e^{\Theta (y)} \notag\\
&=\Phi\left[x^++L/2\right]-\Phi\left[L/2\right],
\end{align}
where the function $\Phi\left[x\right]$ is introduced by
\begin{equation}
\Phi\left[x\right]=\int^{x}_{0}dy\,e^{\Theta (y)}.
\label{eq:Phi}
\end{equation}
The coordinate $x_\text{\three}^-$ is determined by the matching condition:
\begin{align}
x_\text{\three}^{-}\left[x^-\right]&=x_\text{\three}^{+}\left[x^--L\right]+L \notag\\
&=\int^{x^--L}_{L/2} dy\, e^{\Theta (y+L/2)}+L \notag \\
&=\Phi\left[x^--L/2\right]-\Phi\left[L/2\right]+L.
\end{align}
From these equations, the metric in region {\two} can be written as
\small
\begin{align}
ds^2_\text{\two}&=-e^{2\Theta (t)}dx^+dx^- \notag \\
&=-\exp\left[2\Theta(t)-\Theta(x^++L/2)-\Theta(x^--L/2)\right] dx_\text{\three}^+dx_\text{\three}^-.
\end{align}
\normalsize
At the boundary, $x^+=t-L/2$ and $x^-=t+L/2$, so the conformal factor becomes unity. Thus,  the coordinates $x^\pm$ in region {\two} can be extended to  region {\three}.

\subsection{Region I and  {\two}}
Let us consider the matching conditions at $x=x_{\rm{I}}=L/2$ :
\begin{equation}
x_{\rm{I}}^{+}\left[t+L/2\right]-x_{\rm{I}}^{-}\left[t-L/2\right]=L.
\end{equation}
By similar calculation as matching regions {\two} and \three, we obtain
\begin{equation}
x_{\rm{I}}^+\left[x^+\right]
=\int^{x^+}_{0} dy\, e^{\Theta (y-L/2)}
=\Phi\left[x^+-L/2\right]-\Phi\left[-L/2\right],
\end{equation}
and
\begin{align}
x_{\rm{I}}^{-}\left[x^-\right]&=x_{\rm{I}}^{+}\left[x^-+L\right]-L \notag \\
&=\int^{x^-+L}_{0} dy\, e^{\Theta (y-L/2)}-L \notag \\
&=\Phi\left[x^-+L/2\right]-\Phi\left[-L/2\right]-L.
\end{align}
The metric in region {\two} becomes
\small
\begin{align}
ds^2_\text{\two}&=-e^{2\Theta (t)}dx^+dx^- \notag \\
&=-\exp\left[2\Theta(t)-\Theta(x^+-L/2)-\Theta(x^-+L/2)\right] dx_{\rm{I}}^+dx_{\rm{I}}^-.
\end{align}
\normalsize
At the boundary, $x^+=t+L/2$ and $x^-=t-L/2$, so the conformal factor becomes unity, and the coordinates $x^\pm$ can be extended into region I.

\subsection{Region I and {\three}}

Using the function $\Phi\left[x\right]$ introduced above, we can express the relationship between regions I and {\three}. Coordinates $x_{\rm{I}}^{\pm}$ and $x_\text{\three}^{\pm}$ are written as 
\begin{align}
x_\text{\three}^+=&\Phi(x^++L/2)-\Phi(L/2),\\
x_{\rm{I}}^+=&\Phi(x^+-L/2)-\Phi(-L/2), \\
x_\text{\three}^-=&\Phi(x^--L/2)-\Phi(L/2)+L,\\
x_{\rm{I}}^+=&\Phi(x^-+L/2)-\Phi(-L/2)-L.
\end{align}
After eliminating $x^{\pm}$, we obtain relations between $x_\text{I}^\pm$ and $x_\text{\three}^\pm$ as
\begin{align}\label{I}
&x_{\rm{I}}^+=\Phi\left[-L+\Phi^{-1}\left[x_\text{\three}^++\Phi\left[L/2\right]\right]\right]-\Phi\left[-L/2\right], \\
%
&x_\text{\three}^-=\Phi\left[-L+\Phi^{-1}\left[x_{\rm{I}}^-+L+\Phi\left[-L/2\right]\right]\right]-\Phi\left[L/2\right]+L.
\end{align}

\subsection{Example of analog spacetime : de Sitter case}
\begin{figure}[tbh]
\centering
\includegraphics[width=0.9\linewidth]{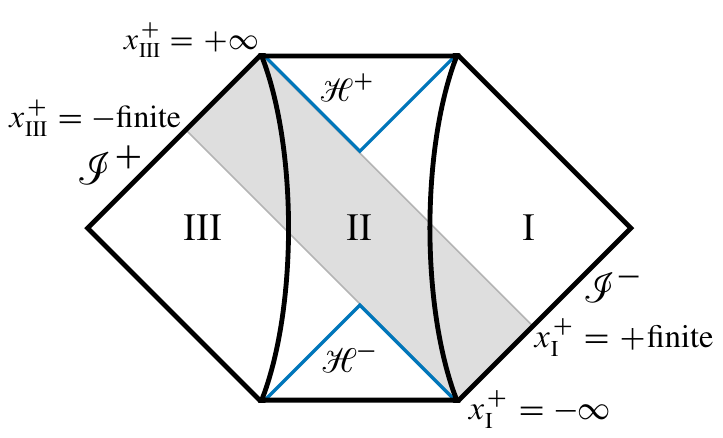}
\caption{Penrose diagram for the QH system with a de Sitter edge region \two. A future horizon and a past horizon exist (time-symmetric eternal case). The vacuum condition is imposed on the past null infinity $\mathcal{I}^-$ in region I.}
\label{fig:fig2}
\end{figure}

For later usage in the analysis of the de Sitter spacetime,
 we calculate the specific form of $\Phi\left[x\right]$ and $\Phi^{-1}\left[x\right]$ (see Fig.~\ref{fig:fig2} for this setup). The conformal factor of the de Sitter spacetime is
\begin{equation}
e^{\Theta(t)}=\frac{1}{\cos (Ht)}.
\end{equation}
\noindent
Substituting this into the definition of $\Phi\left[x\right]$ \eqref{eq:Phi}, we obtain
\begin{align}
&\Phi(x)=\frac{1}{2H}\ln \frac{1+\sin Hx}{1-\sin Hx}, \\
&\Phi^{-1}(x)=\frac{1}{H}\arcsin \tanh Hx .
\end{align}
Then the metric in region {\two{ is written as
\begin{align}
&ds^2_\text{\two}=-\frac{dx_{\rm{I}}^+dx_{\rm{I}}^-}{\cosh^2\left[H(x_{\rm{I}}-L/2)\right]}, \\
&ds^2_\text{\two}=-\frac{dx_\text{\three}^+\,dx_\text{\three}^-}{\cosh^2\left[H(x_\text{\three}+L/2)\right]}.
\end{align}
In the following, we will refer to region {\two} covered by $x_{\rm{I}}$ as A and region covered by $x_\text{\three}$ as B (Fig.~\ref{fig:AB}).
\begin{figure}[bh]
\centering
\includegraphics[width=0.8\linewidth]{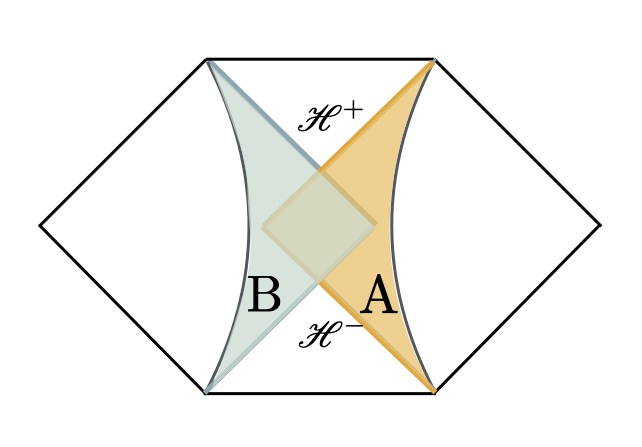}
\caption{Region A in the de Sitter region {\two} is covered by $x_\text{I}^\pm$ and region B in the de Sitter region {\two} is covered by $x_\text{\three}^\pm$.}
\label{fig:AB}
\end{figure}

\section{Application of Anomaly method}
This QH system reproduces a left-moving chiral field in de Sitter spacetime. It is known that in such a situation, not only a trace anomaly but also a gravitational anomaly exists. In $(1+1)$-dimensional spacetimes, they are represented by the following expression which indicates a violation of the conservation law for the expectation value of EMT \cite{Alvarez-Gaume:1983ihn,Bertlmann:1996xk,Bertlmann:2000da}:
\begin{align}
&T^{\alpha}_{\alpha}=\frac{R}{48\pi},\\
&\nabla_{\mu}T^{\mu}_{\nu}=\frac{\epsilon_{\mu\nu}\partial^{\mu}R}{96\pi\sqrt{-g}},
\label{eq:anomaly}
\end{align}
where $R$ denotes the spacetime scalar curvature and $\epsilon_{\mu\nu}$ is the totally antisymmetric tensor in two dimensions.
By solving the anomaly equation \eqref{eq:anomaly}, we can observe the emergence of Hawking radiation in region \text{\three}. 

\subsection{General solution of  anomaly equation}
First, we determine the general solution of the EMT in each region. The scalar curvature of this system is given by 
\begin{equation}
R=2H^2\theta (-x_{\rm{I}}+L/2)\theta (x_\text{\three}+L/2)
\end{equation}
where the unit step function  defined by
\begin{equation}
\theta(x)=\begin{cases}
0&\quad\text{for}\quad x<0,\\
1&\quad\text{for}\quad x\ge 0.
\end{cases}
\end{equation}
The EMT is conserved in bulk regions beside boundaries. Thus in regions I and {\three} (Minkowski region), we solve 
\begin{equation}
{T_i}_{\alpha}^\alpha=0,\quad\partial_{\mu}{T_i}^{\mu}_{\nu}=0,\quad
(i=\text{I, \three}).
\label{eq:Teq0}
\end{equation}
We want to consider the general solution of Eq.~\eqref{eq:Teq0}. By using $T^+_+=T^-_-=1/2(T^t_t+T^x_x)=1/2\,T^{\alpha}_{\alpha}$, Eq.~\eqref{eq:Teq0} becomes
\begin{equation}
\partial_{-}{T_i}^{-}_{+}=0,\quad\partial_{+}{T_i}^{+}_{-}=0.
\end{equation}
Thus, we obtain
\begin{equation}
{T_i}^{-}_{+}=-2{T_i}_{++}=F_i\left[x^+_i\right],\quad {T_i}^{+}_{-}=-2{T_i}_{--}=G_i\left[x^-_i\right],
\label{eq:sol0}
\end{equation}
where $F_i\left[x^+_i\right]$ is an arbitrary function of $x^+_i$ and $G_i\left[x^-_i\right]$ is an arbitrary function of $x^-_i$.

In regions A and B (de Sitter region), the EMT obeys
\begin{equation}
{T_j}_{\alpha}^\alpha=\frac{R}{48\pi},\quad\nabla_{\mu}{T_j}^{\mu}_{\nu}=\frac{\epsilon_{\mu\nu}\partial^{\mu}R}{96\pi\sqrt{-g_j}},\quad
(j=A, B).
\end{equation}
The non-zero Christoffel symbols are ${\Gamma_j}^+_{++}=h_j\partial_+(1/h_j)$ and ${\Gamma_j}^-_{--}=h_j\partial_-(1/h_j)$ with $h_A:=\cosh^2\left[H(x_{\rm{I}}-L/2)\right]$ and $h_B:=\cosh^2\left[H(x_\text{\three}+L/2)\right]$. Therefore, we obtain the following explicit form of the anomaly equation:
\begin{equation}
\begin{split}
\partial_-{T_A}^-_++{\Gamma_A}^-_{--}{T_A}^-_+&=0,\\   
\partial_+{T_A}^+_-+{\Gamma_A}^+_{++}{T_A}^+_-&=-\frac{H^2}{48\pi}\delta (-x_{\rm{I}}+L/2) . 
\end{split}
\end{equation}
\begin{equation}
\begin{split}
\partial_-{T_B}^-_++{\Gamma_B}^-_{--}{T_B}^-_+&=0,\\   
\partial_+{T_B}^+_-+{\Gamma_B}^+_{++}{T_B}^+_-&=\frac{H^2}{48\pi}\delta (x_\text{\three}-L/2) . 
\end{split}
\end{equation}
The solution of these equations are given by
\begin{align}
{T_A}^-_+&=-2\cosh^2\left[H(x_{\rm{I}}-L/2)\right]{T_A}_{++} \notag\\
&=f_A\left[x^+_{\rm{I}}\right]\cosh^2\left[H(x_{\rm{I}}-L/2)\right],
\label{eq:fA}\\   
{T_A}^+_-&=-2\cosh^2\left[H(x_{\rm{I}}-L/2)\right]{T_A}_{--} \notag\\
&=\left[\frac{H^2}{24\pi}\theta(-x_{\rm{I}}+L/2) + a\right] \notag \\
&\qquad\times g_A\left[x^-_{\rm{I}}\right]\cosh^2\left[H(x_{\rm{I}}-L/2)\right],
\label{eq:solA}
\end{align}
\begin{align}
{T_B}^-_+&= -2\cosh^2\left[H(x_\text{\three}+L/2)\right]{T_B}_{++} \notag\\
&=f_B\left[x^+_\text{\three}\right]\cosh^2\left[H(x_\text{\three}+L/2)\right],\\   
{T_B}^+_-&= -2\cosh^2\left[H(x_\text{\three}+L/2)\right]{T_B}_{--}\notag\\
&=\left[\frac{H^2}{24\pi}\theta(x_\text{\three}+L/2) + b\right]\notag \\
&\qquad \times g_B\left[x^-_\text{\three}\right]\cosh^2\left[H(x_\text{\three}+L/2)\right],
\label{eq:solB}
\end{align}
where $a, b$ are integration constants, and $f_j\left[x^+_j\right]$ are arbitrary functions of $x^+_j$, $g_j\left[x^-_j\right]$ are arbitrary functions of $x^-_j$.

In the later section, we will calculate ${T_\text{\three}}_{++}$ as the left-moving Hawking radiation. As this quantity depends only on the coordinate $x_\text{\three}^+$, it is determined by the initial condition in region I and the boundary conditions at $x=\pm L/2$, which are provided by the anomaly equation. 

\subsection{Boundary conditions}
We write the total EMT in region I$\,\cup\,$A as
\begin{equation}
  {T_{\rm{Itot}}}^{\mu}_{\nu}={T_\text{I}}^{\mu}_{\nu}\left[1-\theta(-x_{\rm{I}}+L/2)\right]+{T_{\rm{A}}}^{\mu}_{\nu}\,\theta (-x_{\rm{I}}+L/2).
\end{equation}
Under the diffeomorphism transformation, the change of effective action $W$ is
\begin{equation}
-\delta_{\lambda}W=\int d^2x\sqrt{-g}\,\lambda^{\nu}\nabla_{\mu} {T_{\rm{Itot}}}^{\mu}_{\nu},
\end{equation}
where $\lambda^{\nu}$ are arbitrary variational parameters. To restore the diffeomorphism covariance, each coefficients of $\lambda^{\nu}$ have to vanish. Each coefficient is divergence of the EMT and calculated as
\begin{align}
\nabla_{\mu} {T_{\rm{Itot}}}^{\mu}_{+}&=\nabla_{\mu}{T_\text{I}}^{\mu}_{+}\left[1-\theta (-x_{\rm{I}}+L/2)\right] \notag \\
&\quad+\nabla_{\mu}{T_{\rm{A}}}^{\mu}_{+}\,\theta (-x_{\rm{I}}+L/2) \notag \\
&\quad+({T_{\rm{A}}}^{\mu}_{+}-{T_\text{I}}^{\mu}_{+})\partial_{\mu}\theta (-x_{\rm{I}}+L/2) \notag\\
&=\left(-\frac{H^2}{48\pi}+\frac{1}{2}{T_{\rm{A}}}^{-}_{+}-\frac{1}{2}{T_{\rm{I}}}^{-}_{+}\right)\delta(-x_{\rm{I}}+L/2),
\end{align}
and
\begin{align}
\nabla_{\mu} {T_{\rm{Itot}}}^{\mu}_{-}&=\nabla_{\mu}{T_\text{I}}^{\mu}_{-}\left[1-\theta (-x_{\rm{I}}+L/2)\right] \notag \\
&\quad+\nabla_{\mu}{T_{\rm{A}}}^{\mu}_{-}\,\theta (-x_{\rm{I}}+L/2) \notag \\
&\quad+({T_{\rm{A}}}^{\mu}_{-}-{T_\text{I}}^{\mu}_{-})\partial_{\mu}\theta (-x_{\rm{I}}+L/2) \notag \\
&=\left(-\frac{1}{2}{T_{\rm{A}}}^{+}_{-}+\frac{1}{2}{T_{\rm{I}}}^{+}_{-}\right)\delta(-x_{\rm{I}}+L/2).
\end{align}
To make the total EMT anomaly free, the coefficient of the delta function must  vanish and boundary conditions for $T_{++}$, $T_{--}$  at $x=L/2$ become
\begin{equation}
\begin{split}
{T_{\rm{A}}}_{++}-{T_{\rm{I}}}_{++} &=-\frac{H^2}{48\pi},\\
{T_{\rm{A}}}_{--}-{T_{\rm{I}}}_{--} &=0,
\label{eq:junction1}
\end{split}
\end{equation}
By performing the same calculation in region {\three}$\,\cup\,$B, we obtain following condition at $x=-L/2$:
\begin{equation}
\begin{split}
{T_\text{\three}}_{++}-{T_B}_{++} &=\frac{H^2}{48\pi},\\
{T_\text{\three}}_{--}-{T_B}_{--} &=0.
\label{eq:junction2}
\end{split}
\end{equation}

\subsection{Calculation of Hawking radiation}
From the general solution of the EMT obtained above and the boundary conditions, we can calculate the flux in region {\three}. Since \eqref{eq:sol0}, \eqref{eq:solA} and \eqref{eq:solB} show that $T_{++}$ in each region does not depend on $x^-$, $T_{++}$ in each region is determined once its value is specified at a certain $x^+$. We impose the in-vacuum condition $T_{++}=0$ at $\mathcal{I}^-$ in region I. Then, one finds 
\begin{equation}
{T_{\rm{I}}}_{++}=-\frac{1}{2}F_{\rm{I}}\left[x^+_{\rm{I}}\right]=0.
\end{equation}
With the EMT in region I, using the matching condition \eqref{eq:junction1}, we obtain
\begin{equation}
{T_{\rm{A}}}_{++}={T_{\rm{I}}}_{++}-\frac{H^2}{48\pi}=-\frac{H^2}{48\pi}. 
\end{equation}
This boundary condition is same as to fix the arbitary function $f_A[x_\text{I}^+]$ in Eq.~\eqref{eq:fA} as $f_A[x_\text{I}^+]=H^2/(24\pi)$.

Next, to determine the relationship between ${T_{\rm{A}}}_{++}$ and ${T_{\rm{B}}}_{++}$, we consider a coordinate transformation.
In the overlaped region A$\cap$B, they are related by the coordinate transformation
\begin{equation}
T_{B++}=\left(\frac{\partial x_{\rm{I}}^+}{\partial x_\text{\three}^+}\right)^2T_{A++}.
\end{equation}
Using the formula \eqref{I} ,
\begin{align}
\frac{\partial x_{\rm{I}}^{+}}{\partial x_\text{\three}^{+}}=&\csc{\left[HL+\arccos{\left[\frac{\cosh{H x_\text{\three}^{+}}\sin {\frac{HL}{2}}+\sinh{Hx_\text{\three}^{+}}}{\cosh{H x_\text{\three}^{+}}+\sin {\frac{HL}{2}}\sinh{Hx_\text{\three}^{+}}}\right]}\right]} \notag\\
&\times\sqrt{\frac{\cos^2{\frac{HL}{2}}}{(\cosh{Hx_\text{\three}^{+}}+\sin{\frac{HL}{2}}\sinh{Hx_\text{\three}^{+}})^2}}~,
\end{align}
and ${T_{\rm{B}}}_{++}$ is determined as
\footnotesize
\begin{align}
{T_{\rm{B}}}_{++}
&=-\frac{H^2}{48\pi}\csc^2{\left[HL+\arccos{\left[\frac{\cosh{H x_\text{\three}^{+}}\sin {\frac{HL}{2}}+\sinh{Hx_\text{\three}^{+}}}{\cosh{H x_\text{\three}^{+}}+\sin {\frac{HL}{2}}\sinh{Hx_\text{\three}^{+}}}\right]}\right]} \notag \\
&\quad \times\frac{\cos^2{\frac{HL}{2}}}{(\cosh{Hx_\text{\three}^{+}}+\sin{\frac{HL}{2}}\sinh{Hx_\text{\three}^{+}})^2}.
\end{align}
\normalsize
Finally, by using the boundary condition at $x_\text{\three}=-L/2$ \eqref{eq:junction2}, we obtain the flux in region {\three} as a function of $x_\text{\three}^+$:
\footnotesize
\begin{align}    
{T_{\rm{\three}}}_{++}
&=-\frac{H^2}{48\pi}\csc^2{\left[HL+\arccos{\left[\frac{\cosh{H x_\text{\three}^{+}}\sin {\frac{HL}{2}}+\sinh{Hx_\text{\three}^{+}}}{\cosh{H x_\text{\three}^{+}}+\sin {\frac{HL}{2}}\sinh{Hx_\text{\three}^{+}}}\right]}\right]} \notag\\
&\quad\times\frac{\cos^2{\frac{HL}{2}}}{(\cosh{Hx_\text{\three}^{+}}+\sin{\frac{HL}{2}}\sinh{Hx_\text{\three}^{+}})^2}+\frac{H^2}{48\pi}.
\label{eq:flux-main}
\end{align}
\normalsize
\begin{figure}[th]
\centering
\includegraphics[width=1.0\linewidth]{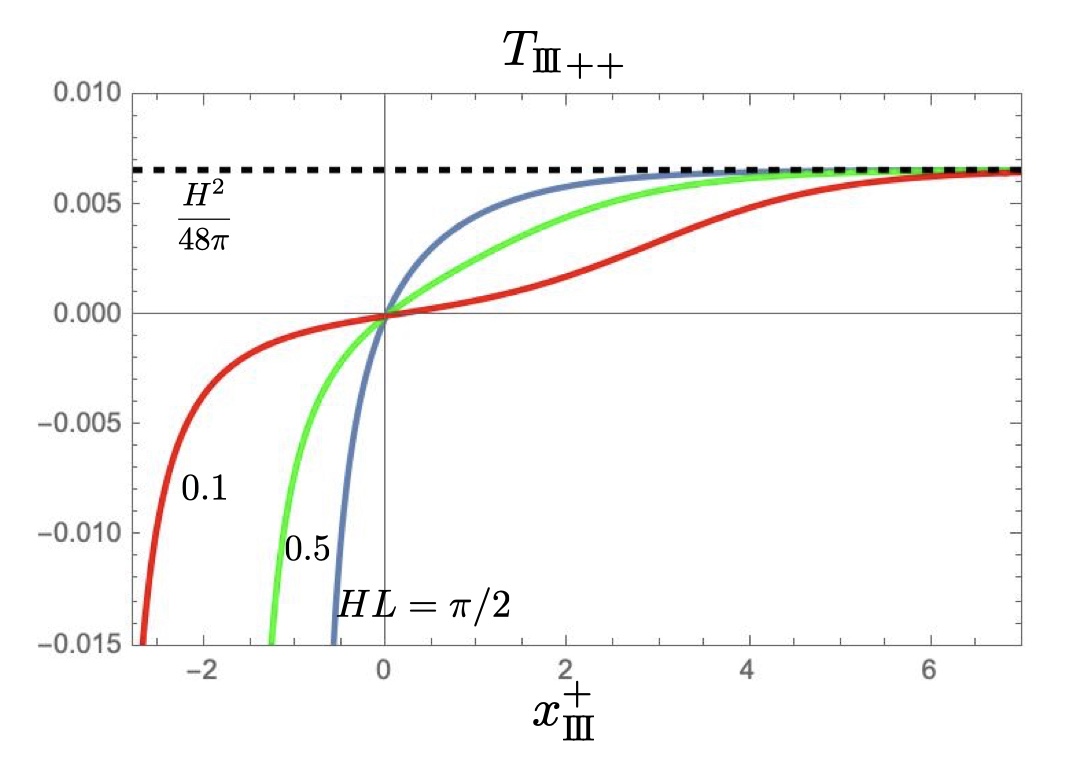}
\caption{$x_\text{\three}^+$ dependence of ${T_{\rm{\three}}}_{++}$ in region {\three} with different values of $L$ $(H=1)$. For $x_\text{\three}^+\rightarrow\infty$, ${T_{\rm{\three}}}_{++}$ approaches to $H^2/(48\pi)$ which corresponds to the outgoing thermal flux with the Gibbons-Hawking temperature $T_\text{H}=H/(2\pi)$. For $x_\text{\three}^+<0$, the left-moving flux is negative which represents negative energy density. For a finite negative $x_\text{\three}^+$, the flux negatively diverges.}
\label{fig:flux}
\end{figure}
 Figure~\ref{fig:flux} shows ${T_{\rm{\three}}}_{++}$ as a function of  $x_{\rm{\three}}^+$. For $x_\text{\three}^+>0$, ${T_{\rm{\three}}}_{++}>0$ and this represents existence of the left propagating positive energy flux.
${T_{\rm{\three}}}_{++}$ approaches to $H^2/(48\pi)$ as $x_\text{\three}^+\rightarrow+\infty$; If we assume thermality of the flux, it reproduces Gibbons-Hawking temperature $T_\text{H}=H/(2\pi)$ which is associated with the future event horizon (black hole horizon) in de Sitter region {\two}. On the other hand, for  $x_\text{\three}^+<0$, the left propagating flux becomes negative and diverges at a finite negative value of  $x_\text{\three}^+$. This behavior of the EMT reflects existence of the negative energy flux associated with the past event horizon $\mathcal{H^{-}}$ (white hole horizon) in region {\two} (see Fig.~\ref{fig:fig2}). 

The flux formula \eqref{eq:flux-main} obtained in this study is correct for the parameter range $LH<\pi/2$. Otherwise, the vacuum condition imposed on $\mathcal{I}^{-}$ in region I does not determine the state in region {\three}. Actually, in such a case, regions A and B do not have overlap and matching procedure fails. As the limiting case $LH\rightarrow 0$, the formula \eqref{eq:flux-main} predicts ${T_{\rm{\three}}}_{++}\rightarrow 0$ and consistently reproduces the Minkowski case with no particle creations.

\section{Summary and Conclusion}
In this paper, Hawking radiation in an analog de Sitter spacetime in the quantum Hall system was calculated by obtaining the general solution of the anomalous conservation law in each region and considering the boundary conditions between region I and {\two}, and between region {\two} and {\three}. In this system, the effect of gravitational anomaly does not appear in the conservation law of the EMT in the bulk regions, but the effect of anomaly appears as the boundary conditions between bulk regions.
Furthermore, in the original calculation of Hawking radiation using gravitational anomaly, the assumption of the existence of anomaly was necessary. However, in the QH system, the calculation becomes exact because of the inherent property that only chiral fields that exist.
\begin{figure}[th]
\centering
\includegraphics[width=0.7\linewidth]{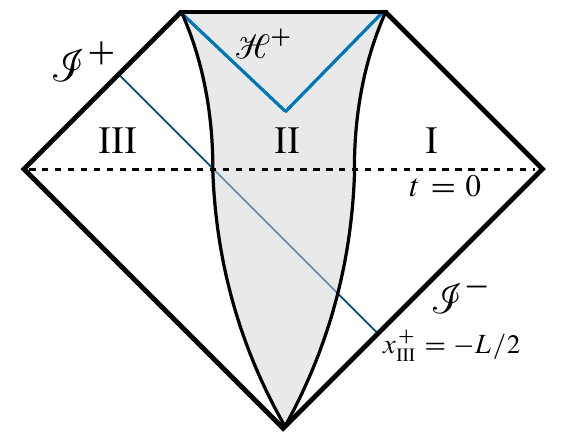}
\caption{Penrose diagram for non-etnernal case that corresponds to black hole formation via gravitational collapse.}
\label{fig:non-eternal}
\end{figure}

It has been found that Hawking radiation can be obtained by calculating gravitational anomaly, but there are several points needed to be discussed. We found a negatively divergent behavior of ${T_\text{\three}}_{++}$ at a negative finite value of $x_\text{\three}^+$. This divergent behavior of the EMT is related to the existence of the past horizon $\mathcal{H}^-$ in region {\two}, but detailed investigation of this issue has not been done.
We assumed that the analog de Sitter spacetime is time-symmetric, which makes boundary conditions simple. However, in order to extract only the effect of Hawking radiation from the future horizon, an analysis using the non-eternal setup~\cite{Nambu2023a}  (Fig.~\ref{fig:non-eternal}) is required. We left this for our next work.

\begin{acknowledgements}
The authors thank Yuuki Sugiyama for valuable discussion on this subject. YN was supported by JSPS KAKENHI (Grant No.~JP22H05257) and MEXT KAKENHI Grant-in-Aid for Transformative Research Areas A ``Extreme Universe"(Grant No.~24H00956).
\end{acknowledgements}

\bibliography{bib-anomaly} 

\begin{thebibliography}{17}%
\makeatletter
\providecommand \@ifxundefined [1]{%
 \@ifx{#1\undefined}
}%
\providecommand \@ifnum [1]{%
 \ifnum #1\expandafter \@firstoftwo
 \else \expandafter \@secondoftwo
 \fi
}%
\providecommand \@ifx [1]{%
 \ifx #1\expandafter \@firstoftwo
 \else \expandafter \@secondoftwo
 \fi
}%
\providecommand \natexlab [1]{#1}%
\providecommand \enquote  [1]{``#1''}%
\providecommand \bibnamefont  [1]{#1}%
\providecommand \bibfnamefont [1]{#1}%
\providecommand \citenamefont [1]{#1}%
\providecommand \href@noop [0]{\@secondoftwo}%
\providecommand \href [0]{\begingroup \@sanitize@url \@href}%
\providecommand \@href[1]{\@@startlink{#1}\@@href}%
\providecommand \@@href[1]{\endgroup#1\@@endlink}%
\providecommand \@sanitize@url [0]{\catcode `\\12\catcode `\$12\catcode `\&12\catcode `\#12\catcode `\^12\catcode `\_12\catcode `\%12\relax}%
\providecommand \@@startlink[1]{}%
\providecommand \@@endlink[0]{}%
\providecommand \url  [0]{\begingroup\@sanitize@url \@url }%
\providecommand \@url [1]{\endgroup\@href {#1}{\urlprefix }}%
\providecommand \urlprefix  [0]{URL }%
\providecommand \Eprint [0]{\href }%
\providecommand \doibase [0]{https://doi.org/}%
\providecommand \selectlanguage [0]{\@gobble}%
\providecommand \bibinfo  [0]{\@secondoftwo}%
\providecommand \bibfield  [0]{\@secondoftwo}%
\providecommand \translation [1]{[#1]}%
\providecommand \BibitemOpen [0]{}%
\providecommand \bibitemStop [0]{}%
\providecommand \bibitemNoStop [0]{.\EOS\space}%
\providecommand \EOS [0]{\spacefactor3000\relax}%
\providecommand \BibitemShut  [1]{\csname bibitem#1\endcsname}%
\let\auto@bib@innerbib\@empty
\bibitem [{\citenamefont {Hotta}\ \emph {et~al.}(2022)\citenamefont {Hotta}, \citenamefont {Nambu}, \citenamefont {Sugiyama}, \citenamefont {Yamamoto},\ and\ \citenamefont {Yusa}}]{Hotta2022b}%
  \BibitemOpen
  \bibfield  {author} {\bibinfo {author} {\bibfnamefont {M.}~\bibnamefont {Hotta}}, \bibinfo {author} {\bibfnamefont {Y.}~\bibnamefont {Nambu}}, \bibinfo {author} {\bibfnamefont {Y.}~\bibnamefont {Sugiyama}}, \bibinfo {author} {\bibfnamefont {K.}~\bibnamefont {Yamamoto}},\ and\ \bibinfo {author} {\bibfnamefont {G.}~\bibnamefont {Yusa}},\ }\bibfield  {title} {\bibinfo {title} {{Expanding edges of quantum Hall systems in a cosmology language: Hawking radiation from de Sitter horizon in edge modes}},\ }\href {https://doi.org/10.1103/PhysRevD.105.105009} {\bibfield  {journal} {\bibinfo  {journal} {Phys. Rev. D}\ }\textbf {\bibinfo {volume} {105}},\ \bibinfo {pages} {105009} (\bibinfo {year} {2022})},\ \Eprint {https://arxiv.org/abs/2202.03731} {arXiv:2202.03731} \BibitemShut {NoStop}%
\bibitem [{\citenamefont {Nambu}\ and\ \citenamefont {Hotta}(2023)}]{Nambu2023a}%
  \BibitemOpen
  \bibfield  {author} {\bibinfo {author} {\bibfnamefont {Y.}~\bibnamefont {Nambu}}\ and\ \bibinfo {author} {\bibfnamefont {M.}~\bibnamefont {Hotta}},\ }\bibfield  {title} {\bibinfo {title} {{Analog de Sitter universe in quantum Hall systems with an expanding edge}},\ }\href {https://doi.org/10.1103/PhysRevD.107.085002} {\bibfield  {journal} {\bibinfo  {journal} {Phys. Rev. D}\ }\textbf {\bibinfo {volume} {107}},\ \bibinfo {pages} {085002} (\bibinfo {year} {2023})},\ \Eprint {https://arxiv.org/abs/2301.09270} {arXiv:2301.09270} \BibitemShut {NoStop}%
\bibitem [{\citenamefont {Alvarez-Gaume}\ and\ \citenamefont {Witten}(1984)}]{Alvarez-Gaume:1983ihn}%
  \BibitemOpen
  \bibfield  {author} {\bibinfo {author} {\bibfnamefont {L.}~\bibnamefont {Alvarez-Gaume}}\ and\ \bibinfo {author} {\bibfnamefont {E.}~\bibnamefont {Witten}},\ }\bibfield  {title} {\bibinfo {title} {{Gravitational Anomalies}},\ }\href {https://doi.org/10.1016/0550-3213(84)90066-X} {\bibfield  {journal} {\bibinfo  {journal} {Nucl. Phys. B}\ }\textbf {\bibinfo {volume} {234}},\ \bibinfo {pages} {269} (\bibinfo {year} {1984})}\BibitemShut {NoStop}%
\bibitem [{\citenamefont {Bertlmann}(1996)}]{Bertlmann:1996xk}%
  \BibitemOpen
  \bibfield  {author} {\bibinfo {author} {\bibfnamefont {R.~A.}\ \bibnamefont {Bertlmann}},\ }\href@noop {} {\emph {\bibinfo {title} {{Anomalies in quantum field theory}}}}\ (\bibinfo {year} {1996})\BibitemShut {NoStop}%
\bibitem [{\citenamefont {Bertlmann}\ and\ \citenamefont {Kohlprath}(2001)}]{Bertlmann:2000da}%
  \BibitemOpen
  \bibfield  {author} {\bibinfo {author} {\bibfnamefont {R.~A.}\ \bibnamefont {Bertlmann}}\ and\ \bibinfo {author} {\bibfnamefont {E.}~\bibnamefont {Kohlprath}},\ }\bibfield  {title} {\bibinfo {title} {{Two-dimensional gravitational anomalies, Schwinger terms and dispersion relations}},\ }\href {https://doi.org/10.1006/aphy.2000.6110} {\bibfield  {journal} {\bibinfo  {journal} {Annals Phys.}\ }\textbf {\bibinfo {volume} {288}},\ \bibinfo {pages} {137} (\bibinfo {year} {2001})},\ \Eprint {https://arxiv.org/abs/hep-th/0011067} {arXiv:hep-th/0011067} \BibitemShut {NoStop}%
\bibitem [{\citenamefont {Hawking}(1975)}]{Hawking:1975vcx}%
  \BibitemOpen
  \bibfield  {author} {\bibinfo {author} {\bibfnamefont {S.~W.}\ \bibnamefont {Hawking}},\ }\bibfield  {title} {\bibinfo {title} {{Particle Creation by Black Holes}},\ }\href {https://doi.org/10.1007/BF02345020} {\bibfield  {journal} {\bibinfo  {journal} {Commun. Math. Phys.}\ }\textbf {\bibinfo {volume} {43}},\ \bibinfo {pages} {199} (\bibinfo {year} {1975})},\ \bibinfo {note} {[Erratum: Commun.Math.Phys. 46, 206 (1976)]}\BibitemShut {NoStop}%
\bibitem [{\citenamefont {Gibbons}\ and\ \citenamefont {Hawking}(1977)}]{Gibbons:1976ue}%
  \BibitemOpen
  \bibfield  {author} {\bibinfo {author} {\bibfnamefont {G.~W.}\ \bibnamefont {Gibbons}}\ and\ \bibinfo {author} {\bibfnamefont {S.~W.}\ \bibnamefont {Hawking}},\ }\bibfield  {title} {\bibinfo {title} {{Action Integrals and Partition Functions in Quantum Gravity}},\ }\href {https://doi.org/10.1103/PhysRevD.15.2752} {\bibfield  {journal} {\bibinfo  {journal} {Phys. Rev. D}\ }\textbf {\bibinfo {volume} {15}},\ \bibinfo {pages} {2752} (\bibinfo {year} {1977})}\BibitemShut {NoStop}%
\bibitem [{\citenamefont {Christensen}\ and\ \citenamefont {Fulling}(1977)}]{Christensen:1977jc}%
  \BibitemOpen
  \bibfield  {author} {\bibinfo {author} {\bibfnamefont {S.~M.}\ \bibnamefont {Christensen}}\ and\ \bibinfo {author} {\bibfnamefont {S.~A.}\ \bibnamefont {Fulling}},\ }\bibfield  {title} {\bibinfo {title} {{Trace Anomalies and the Hawking Effect}},\ }\href {https://doi.org/10.1103/PhysRevD.15.2088} {\bibfield  {journal} {\bibinfo  {journal} {Phys. Rev. D}\ }\textbf {\bibinfo {volume} {15}},\ \bibinfo {pages} {2088} (\bibinfo {year} {1977})}\BibitemShut {NoStop}%
\bibitem [{\citenamefont {Birrell}\ and\ \citenamefont {Davies}(1984)}]{Birrell1984}%
  \BibitemOpen
  \bibfield  {author} {\bibinfo {author} {\bibfnamefont {N.~D.}\ \bibnamefont {Birrell}}\ and\ \bibinfo {author} {\bibfnamefont {P.~C.~W.}\ \bibnamefont {Davies}},\ }\href@noop {} {\emph {\bibinfo {title} {{Quantum fields in curved space}}}}\ (\bibinfo  {publisher} {Cambridge University Press},\ \bibinfo {year} {1984})\BibitemShut {NoStop}%
\bibitem [{\citenamefont {Parikh}\ and\ \citenamefont {Wilczek}(2000)}]{Parikh:1999mf}%
  \BibitemOpen
  \bibfield  {author} {\bibinfo {author} {\bibfnamefont {M.~K.}\ \bibnamefont {Parikh}}\ and\ \bibinfo {author} {\bibfnamefont {F.}~\bibnamefont {Wilczek}},\ }\bibfield  {title} {\bibinfo {title} {{Hawking radiation as tunneling}},\ }\href {https://doi.org/10.1103/PhysRevLett.85.5042} {\bibfield  {journal} {\bibinfo  {journal} {Phys. Rev. Lett.}\ }\textbf {\bibinfo {volume} {85}},\ \bibinfo {pages} {5042} (\bibinfo {year} {2000})},\ \Eprint {https://arxiv.org/abs/hep-th/9907001} {arXiv:hep-th/9907001} \BibitemShut {NoStop}%
\bibitem [{\citenamefont {Robinson}\ and\ \citenamefont {Wilczek}(2005)}]{Robinson:2005pd}%
  \BibitemOpen
  \bibfield  {author} {\bibinfo {author} {\bibfnamefont {S.~P.}\ \bibnamefont {Robinson}}\ and\ \bibinfo {author} {\bibfnamefont {F.}~\bibnamefont {Wilczek}},\ }\bibfield  {title} {\bibinfo {title} {{A Relationship between Hawking radiation and gravitational anomalies}},\ }\href {https://doi.org/10.1103/PhysRevLett.95.011303} {\bibfield  {journal} {\bibinfo  {journal} {Phys. Rev. Lett.}\ }\textbf {\bibinfo {volume} {95}},\ \bibinfo {pages} {011303} (\bibinfo {year} {2005})},\ \Eprint {https://arxiv.org/abs/gr-qc/0502074} {arXiv:gr-qc/0502074} \BibitemShut {NoStop}%
\bibitem [{\citenamefont {Iso}\ \emph {et~al.}(2006{\natexlab{a}})\citenamefont {Iso}, \citenamefont {Umetsu},\ and\ \citenamefont {Wilczek}}]{Iso:2006ut}%
  \BibitemOpen
  \bibfield  {author} {\bibinfo {author} {\bibfnamefont {S.}~\bibnamefont {Iso}}, \bibinfo {author} {\bibfnamefont {H.}~\bibnamefont {Umetsu}},\ and\ \bibinfo {author} {\bibfnamefont {F.}~\bibnamefont {Wilczek}},\ }\bibfield  {title} {\bibinfo {title} {{Anomalies, Hawking radiations and regularity in rotating black holes}},\ }\href {https://doi.org/10.1103/PhysRevD.74.044017} {\bibfield  {journal} {\bibinfo  {journal} {Phys. Rev. D}\ }\textbf {\bibinfo {volume} {74}},\ \bibinfo {pages} {044017} (\bibinfo {year} {2006}{\natexlab{a}})},\ \Eprint {https://arxiv.org/abs/hep-th/0606018} {arXiv:hep-th/0606018} \BibitemShut {NoStop}%
\bibitem [{\citenamefont {Iso}\ \emph {et~al.}(2006{\natexlab{b}})\citenamefont {Iso}, \citenamefont {Umetsu},\ and\ \citenamefont {Wilczek}}]{Iso:2006wa}%
  \BibitemOpen
  \bibfield  {author} {\bibinfo {author} {\bibfnamefont {S.}~\bibnamefont {Iso}}, \bibinfo {author} {\bibfnamefont {H.}~\bibnamefont {Umetsu}},\ and\ \bibinfo {author} {\bibfnamefont {F.}~\bibnamefont {Wilczek}},\ }\bibfield  {title} {\bibinfo {title} {{Hawking radiation from charged black holes via gauge and gravitational anomalies}},\ }\href {https://doi.org/10.1103/PhysRevLett.96.151302} {\bibfield  {journal} {\bibinfo  {journal} {Phys. Rev. Lett.}\ }\textbf {\bibinfo {volume} {96}},\ \bibinfo {pages} {151302} (\bibinfo {year} {2006}{\natexlab{b}})},\ \Eprint {https://arxiv.org/abs/hep-th/0602146} {arXiv:hep-th/0602146} \BibitemShut {NoStop}%
\bibitem [{\citenamefont {Murata}\ and\ \citenamefont {Soda}(2006)}]{Murata:2006pt}%
  \BibitemOpen
  \bibfield  {author} {\bibinfo {author} {\bibfnamefont {K.}~\bibnamefont {Murata}}\ and\ \bibinfo {author} {\bibfnamefont {J.}~\bibnamefont {Soda}},\ }\bibfield  {title} {\bibinfo {title} {{Hawking radiation from rotating black holes and gravitational anomalies}},\ }\href {https://doi.org/10.1103/PhysRevD.74.044018} {\bibfield  {journal} {\bibinfo  {journal} {Phys. Rev. D}\ }\textbf {\bibinfo {volume} {74}},\ \bibinfo {pages} {044018} (\bibinfo {year} {2006})},\ \Eprint {https://arxiv.org/abs/hep-th/0606069} {arXiv:hep-th/0606069} \BibitemShut {NoStop}%
\bibitem [{\citenamefont {Jiang}\ \emph {et~al.}(2007)\citenamefont {Jiang}, \citenamefont {Wu},\ and\ \citenamefont {Cai}}]{Jiang:2007wj}%
  \BibitemOpen
  \bibfield  {author} {\bibinfo {author} {\bibfnamefont {Q.-Q.}\ \bibnamefont {Jiang}}, \bibinfo {author} {\bibfnamefont {S.-Q.}\ \bibnamefont {Wu}},\ and\ \bibinfo {author} {\bibfnamefont {X.}~\bibnamefont {Cai}},\ }\bibfield  {title} {\bibinfo {title} {{Hawking radiation from the dilatonic black holes via anomalies}},\ }\href {https://doi.org/10.1103/PhysRevD.76.029904} {\bibfield  {journal} {\bibinfo  {journal} {Phys. Rev. D}\ }\textbf {\bibinfo {volume} {75}},\ \bibinfo {pages} {064029} (\bibinfo {year} {2007})},\ \bibinfo {note} {[Erratum: Phys.Rev.D 76, 029904 (2007)]},\ \Eprint {https://arxiv.org/abs/hep-th/0701235} {arXiv:hep-th/0701235} \BibitemShut {NoStop}%
\bibitem [{\citenamefont {Akhmedova}\ \emph {et~al.}(2010)\citenamefont {Akhmedova}, \citenamefont {Pilling}, \citenamefont {de~Gill},\ and\ \citenamefont {Singleton}}]{Akhmedova:2010zz}%
  \BibitemOpen
  \bibfield  {author} {\bibinfo {author} {\bibfnamefont {V.~E.}\ \bibnamefont {Akhmedova}}, \bibinfo {author} {\bibfnamefont {T.}~\bibnamefont {Pilling}}, \bibinfo {author} {\bibfnamefont {A.}~\bibnamefont {de~Gill}},\ and\ \bibinfo {author} {\bibfnamefont {D.}~\bibnamefont {Singleton}},\ }\bibfield  {title} {\bibinfo {title} {{Tunneling/WKB and anomaly methods for Rindler and de Sitter space-times}},\ }\href {https://doi.org/10.1007/s11232-010-0061-z} {\bibfield  {journal} {\bibinfo  {journal} {Theor. Math. Phys.}\ }\textbf {\bibinfo {volume} {163}},\ \bibinfo {pages} {774} (\bibinfo {year} {2010})}\BibitemShut {NoStop}%
\bibitem [{\citenamefont {Zampeli}(2013)}]{Zampeli:2013lka}%
  \BibitemOpen
  \bibfield  {author} {\bibinfo {author} {\bibfnamefont {A.}~\bibnamefont {Zampeli}},\ }\bibfield  {title} {\bibinfo {title} {{Radiation from horizons, chirality and the principle of effective theory}},\ }\href {https://doi.org/10.1088/1742-6596/453/1/012024} {\bibfield  {journal} {\bibinfo  {journal} {J. Phys. Conf. Ser.}\ }\textbf {\bibinfo {volume} {453}},\ \bibinfo {pages} {012024} (\bibinfo {year} {2013})}\BibitemShut {NoStop}%
\end{thebibliography}%

\end{document}